\documentclass[preprint,3p]{elsarticle}
\usepackage{dsfont,amsmath}
\biboptions{sort&compress}
\journal{Physics Letters B}

\begin{document}
\begin{frontmatter}
\title{A Higher Order GUP with Minimal Length Uncertainty and Maximal Momentum II: Applications}
\author{Pouria Pedram}
\ead{p.pedram@srbiau.ac.ir}
\address{Department of Physics, Science and Research Branch, Islamic Azad University, Tehran, Iran}

\begin{abstract}
In a recent paper, we presented a nonperturbative higher order
generalized uncertainty principle (GUP) that is consistent with
various proposals of quantum gravity such as string theory, loop
quantum gravity, doubly special relativity, and predicts both a
minimal length uncertainty and a maximal observable momentum. In
this Letter, we find exact maximally localized states and present a
formally self-adjoint and naturally perturbative representation of
this modified algebra. Then we extend this GUP to $D$ dimensions
that will be shown it is noncommutative and find invariant density
of states. We show that the presence of the maximal momentum results
in upper bounds on the energy spectrum of the free particle and the
particle in box. Moreover, this form of GUP modifies blackbody
radiation spectrum at high frequencies and predicts a finite
cosmological constant. Although it does not solve the cosmological
constant problem, it gives a better estimation with respect to the
presence of just the minimal length.
\end{abstract}


\end{frontmatter}

\section{Introduction}
The modification of the Heisenberg uncertainty principle in the
context of the Generalized Uncertainty Principle (GUP) and the
Modified Dispersion Relation (MDR) has attracted much attention in
recent years \cite{felder}. This interest arises from various
theories of quantum gravity such as string theory \cite{1,2,3,4},
loop quantum gravity \cite{5}, noncommutative spacetime
\cite{6,7,8}, and doubly special relativity (DSR) \cite{21,22,23}.
All GUP proposals imply the existence of a minimal length scale of
the order of the Planck length
$\ell_{Pl}=\sqrt{\frac{G\hbar}{c^3}}\approx 10^{-35}m$ where $G$ is
Newton's gravitational constant (see for instance
\cite{14,18,19,101,banerjee,p2,p3,p4,p5,p6,NP,102,103,104,105,106,main,Nouicer,pedramPRD,pedramPLB}).
Moreover, a perturbative GUP proposal that is consistent with DSR
theories is studied in
Refs.~\cite{pedram,pedram1,main1,main2,main3,KE}.

Recently, we have proposed a nonperturbative higher order
generalized uncertainty principle which implies both a minimal
length uncertainty and a maximal observable momentum
\cite{pedramPLBm}
\begin{eqnarray}\label{guph}
[X,P]=\frac{i\hbar}{1-\beta P^2}.
\end{eqnarray}
This commutation relation agrees with Kempf, Mangano and Mann (KMM)
\cite{7} and Noucier's \cite{Nouicer} proposals to the leading order
of the GUP parameter $\beta$. In momentum space, the position and
momentum operators can be written as \cite{pedramPLBm}
\begin{eqnarray}\label{rep1}
P \phi(p)&=& p\,\phi(p),\\
X\phi(p)&=& \frac{i\hbar}{1 - \beta
p^2}\partial_p\phi(p).\label{rep2}
\end{eqnarray}
So the completeness relation and the scalar product take the
following form:
\begin{eqnarray}\label{scalP}
\langle\psi|\phi\rangle&=&\int_{-1/\sqrt{\beta}}^{+1/\sqrt{\beta}}\mathrm{d}p
\left(1-\beta p^2\right)\psi^{*}(p)\phi(p),\\
\langle p|p'\rangle&=& \frac{\delta(p-p')}{1-\beta p^2}.
\end{eqnarray}
Also the momentum of the particle is bounded from above
\begin{eqnarray}
P_{max}=\frac{1}{\sqrt{\beta}},
\end{eqnarray}
and the absolutely smallest uncertainty in position is
\begin{eqnarray}
(\Delta X)_{min}=\frac{3\sqrt{3}}{4}\hbar\sqrt{\beta}.
\end{eqnarray}
Approximate maximally localization states (using KMM approach) and
quantum mechanical and semiclassical solutions of the harmonic
oscillator have been also obtained in this framework
\cite{pedramPLBm}.

Here, we first find maximally localized states using Detournay,
Gabriel and Spindel approach. Then we present a formally
self-adjoint representation and study the problems of the free
particle and the particle in a box and show that their energy
spectrum are bounded from above. We also address the generalization
to $D$ dimensions, validity of semiclassical approximation,
invariant density of states, cosmological constant, and blackbody
radiation in this GUP framework.

\section{Maximally localized states}
In KMM approach the maximally localized states are the solutions of
the following equation \cite{7}
\begin{eqnarray}
\bigg(X-\langle X\rangle+\frac{\langle[X,P]\rangle}{2(\Delta
P)^2}\left(P-\langle P\rangle\right)\bigg)|\psi\rangle=0,
\end{eqnarray}
where $[X,P]=if(P)$. However, unlike the ordinary quantum mechanics
where $f(P)=\mathds{1}\hbar$ and therefore $\langle
f(P)\rangle=\hbar$ for all states, in general, the expectation value
of $[X,P]$ depends on the state considered \cite{ref,pedramPLBm}.
So, except $f(P)\sim1+\beta P^2$, it is impossible, for an arbitrary
function $f(P)$, to write any exact solution for the above equation
(see \cite{pedramPLBm} for an approximate solution). On the other
hand, Detournay and collaborators proposed an alternative general
scheme for finding such states based on a constrained variational
principle \cite{ref}. In this framework, the maximally localized
states are the solutions of the following Euler-Lagrange equation in
momentum space
\begin{equation}
\left[-\left(f(p)\partial_p\right)^2-\xi^2+2a\left(if(p)\partial_p-\xi\right)+
2b\left(v(p)-\gamma\right)-\mu^2\right]\psi(p)=0,
\end{equation}
where $a$ and $b$ are Lagrange multipliers and
\begin{eqnarray}
(\Delta X)^2_{min}={\rm
min}\frac{\langle\psi|X^2-\xi^2|\psi\rangle}{\langle\psi|\psi\rangle}\equiv
\mu^2,\quad\quad\xi
=\frac{\langle\psi|X|\psi\rangle}{\langle\psi|\psi\rangle},\quad\quad
\gamma=\frac{\langle\psi|v(p)|\psi\rangle}{\langle\psi|\psi\rangle}.
\end{eqnarray}
Here $v(p)$ is an arbitrary function whose expectation value is
finite (see \cite{ref} for details). Now if we define
\begin{equation}
z(p)=\int_0^pf^{-1}(q)\,\mathrm{d}q,
\end{equation}
and
\begin{equation}
z\left(+P_{max}\right)=\alpha_+>0,\qquad\qquad
z\left(-P_{max}\right)=\alpha_-<0,
\end{equation}
the normalized solution for $b=0$ is \cite{ref}
\begin{eqnarray}
\psi^{\mathrm{ML}}_{\xi}(p)=C\,\exp[-i\,\xi\,z(p)]\,\sin\left\{\mu\left[z(p)-\alpha_-\right]\right\},
\end{eqnarray}
where
\begin{eqnarray}
|C|=\sqrt{\frac{2/\hbar}{\alpha_+-\alpha_-}},\quad\qquad
\mu=\frac{n\pi}{\alpha_+-\alpha_-},\quad\qquad n\in \mathds{N},
\end{eqnarray}
and the corresponding spread in position is given by
\begin{equation}
\left(\Delta
X\right)_{min}\Big|_{b=0}=\frac{\pi}{\alpha_+-\alpha_-}.
\end{equation}

For our case, i.e.~$f(P)=\hbar/\left(1-\beta P^2\right)$, we obtain
\begin{equation}
z(p)=\hbar^{-1}\left(p-\frac{\beta}{3}p^3\right),
\end{equation}
and
\begin{equation}
\alpha_+=+\frac{2}{3\hbar\sqrt{\beta}},\qquad\qquad
\alpha_-=-\frac{2}{3\hbar\sqrt{\beta}}.
\end{equation}
So the solution is
\begin{eqnarray}\nonumber
\psi^{\mathrm{ML}}_{\xi}(p)&=&\sqrt{\frac{3\sqrt{\beta}}{2}}\,\exp\left[\frac{-i\xi}{\hbar}\,\left(p-\frac{\beta}{3}p^3\right)\right]
\,\sin\left[\frac{\mu}{\hbar}\left(p-\frac{\beta}{3}p^3+\frac{2}{3\sqrt{\beta}}\right)\right],\\
&=&\sqrt{\frac{3\sqrt{\beta}}{2}}\,\exp\left[\frac{-i\xi}{\hbar}\,\left(p-\frac{\beta}{3}p^3\right)\right]
\,\cos\left[\frac{3\pi}{4}\sqrt{\beta}\left(p-\frac{\beta}{3}p^3\right)\right],
\end{eqnarray}
and
\begin{equation}
\left(\Delta
X\right)_{min}\Big|_{b=0}=\frac{3\pi}{4}\hbar\sqrt{\beta}.
\end{equation}
Note that $\left(\Delta X\right)_{min}|_{b=0}$ corresponds to a
(local) minimum with respect to $\gamma$ and
$\psi^{\mathrm{ML}}_{\xi}(p)$ is normalized subject to the scalar
product presented in Eq.~(\ref{scalP}). Also the maximally localized
states are not mutually orthogonal
\begin{eqnarray}\nonumber
\langle\psi^{\mathrm{ML}}_{\xi'}|\psi^{\mathrm{ML}}_{\xi}\rangle&=&\frac{3\sqrt{\beta}}{2}\int_{-1/\sqrt{\beta}}^{+1/\sqrt{\beta}}\mathrm{d}p
\left(1-\beta p^2\right)
\,\exp\left[\frac{-i(\xi-\xi')}{\hbar}\,\left(p-\frac{\beta}{3}p^3\right)\right]
\,\cos^2\left[\frac{3\pi}{4}\sqrt{\beta}\left(p-\frac{\beta}{3}p^3\right)\right],\\\nonumber
&=&\frac{3\sqrt{\beta}}{2}\int_{-\frac{2}{3\sqrt{\beta}}}^{+\frac{2}{3\sqrt{\beta}}}\mathrm{d}z
\,\exp\left[\frac{-i(\xi-\xi')z}{\hbar}\right]
\,\cos^2\left[\frac{3\pi}{4}\sqrt{\beta}z\right],\\
&=&\left[\frac{2(\xi-\xi')}{3\hbar\sqrt{\beta}}-\frac{1}{\pi^2}\left(\frac{2(\xi-\xi')}{3\hbar\sqrt{\beta}}\right)^3\right]^{-1}
\sin\left[\frac{2(\xi-\xi')}{3\hbar\sqrt{\beta}}\right],
\end{eqnarray}
as well as KMM proposal which is due to the fuzziness of space in
both frameworks. Now we can define the quasiposition wave function
as
\begin{eqnarray}\label{psiQP}
\psi_{QP}(\xi)\equiv\langle\psi^{\mathrm{ML}}_{\xi}|\phi\rangle=\sqrt{\frac{3\sqrt{\beta}}{2}}\int_{-1/\sqrt{\beta}}^{+1/\sqrt{\beta}}\mathrm{d}p
\left(1-\beta p^2\right)
\,\exp\left[\frac{i\xi}{\hbar}\,\left(p-\frac{\beta}{3}p^3\right)\right]
\,\cos\left[\frac{3\pi}{4}\sqrt{\beta}\left(p-\frac{\beta}{3}p^3\right)\right]\,\phi(p).
\end{eqnarray}
So the inverse transformation reads
\begin{eqnarray}
\phi(p)=\frac{1}{\sqrt{6\sqrt{\beta}}\pi\hbar}\int_{-\infty}^{+\infty}\mathrm{d}\xi
\,\frac{\exp\left[-\frac{i}{\hbar}\xi\left(p-\frac{\beta}{3}p^3\right)\right]}{\cos\left[\frac{3\pi}{4}\sqrt{\beta}\left(p-\frac{\beta}{3}p^3\right)\right]}
\,\,\psi_{QP}(\xi).
\end{eqnarray}
Moreover, the  scalar  product  of  states  in terms  of
quasiposition  wave functions is given by
\begin{eqnarray}\nonumber
\hspace{-1cm}\langle\psi|\phi\rangle&=&\int_{-1/\sqrt{\beta}}^{+1/\sqrt{\beta}}\mathrm{d}p
\left(1-\beta p^2\right)\psi^{*}(p)\phi(p),\\
&=&
\frac{1}{6\sqrt{\beta}\pi^2\hbar^2}\int_{-\infty}^{+\infty}\int_{-\infty}^{+\infty}\int_{-1/\sqrt{\beta}}^{+1/\sqrt{\beta}}\mathrm{d}p\,\mathrm{d}\xi\,\mathrm{d}\xi'
\frac{\left(1-\beta p^2\right)}
{\cos^2\left[\frac{3\pi\sqrt{\beta}}{4}\left(p-\frac{\beta}{3}p^3\right)\right]}\exp\left[\frac{i}{\hbar}(\xi-\xi')\left(p-\frac{\beta}{3}p^3\right)\right]\psi_{QP}^*(\xi)\psi_{QP}(\xi').\hspace{1cm}
\end{eqnarray}

\section{Formally self-adjoint representation}
\begin{figure}
\begin{center}
\includegraphics[width=8cm]{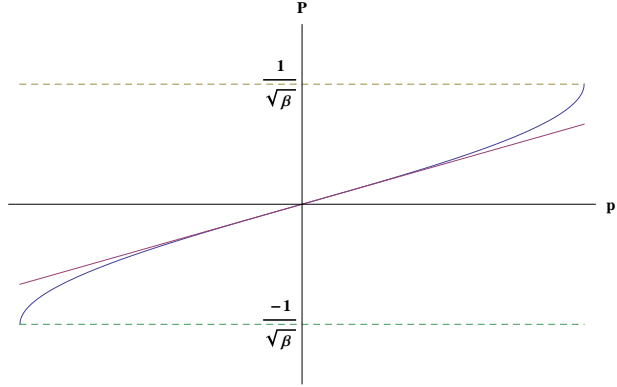}
\caption{\label{fig3}Schematic behavior of $P$ versus $p$ in the
second representation for the ordinary quantum mechanics (red line)
and the GUP framework (blue line).}
\end{center}
\end{figure}

Although the set of Eqs.~(\ref{rep1}) and (\ref{rep2}) is an exact
representation of the algebra presented in Eq.~(\ref{guph}), it does
not preserve the ordinary nature of the position operator.
Alternatively, we can write $P=f(p)$ and retain the ordinary form of
the position operator, i.e., $X=x$ where $[x,p]=i\hbar$. Thus, using
Eq.~(\ref{guph}) we find $\displaystyle
\frac{df}{dp}=\frac{1}{1-\beta f^2}$ which results in
\begin{eqnarray}\label{Pp}
f(p)-\frac{1}{3}\beta f^3(p)=p.
\end{eqnarray}
Consequently, the alternative representation in exact and
perturbative forms is
\begin{eqnarray}\label{PP0}
X &=& x,\\
P&=&
\frac{1-i\sqrt{3}+(-2\beta)^{1/3}\left(3p+\sqrt{9p^2-4/\beta}\right)^{2/3}}{(2\beta)^{2/3}\left(3p+\sqrt{9p^2-4/\beta}\right)^{1/3}},\label{PP}\\
&=&p+\frac{1}{3}\beta p^3+\frac{1}{3}\beta^2 p^5+\frac{4}{9}\beta^3
p^7+\cdots\,.\label{PP2}
\end{eqnarray}
Note that this representation is formally self-adjoint, i.e.,
$A=A^{\dagger}$ for $A\in\{X,P\}$. Also, the presence of the maximal
momentum $P_{max}=1/\sqrt{\beta}$ is manifest from Eq.~(\ref{PP})
which occurs at $p=\displaystyle\frac{2}{3\sqrt{\beta}}$. Now $X$
and $P$ are symmetric operator on the dense domain $S_{\infty}$ with
respect to the following scalar product in the momentum space:
\begin{eqnarray}
\langle\psi|\phi\rangle=\int_{-\frac{2}{3\sqrt{\beta}}}^{+\frac{2}{3\sqrt{\beta}}}\psi^{*}(p)\phi(p)\,\mathrm{d}p.
\end{eqnarray}
We have schematically depicted the behavior of
$P$ versus $p$ in Fig~\ref{fig3}.

In this representation, to write the Hamiltonian, it is more
appropriate to use Eq.~(\ref{PP2}) and express the Hamiltonian
perturbatively as
\begin{eqnarray}
H=\frac{p^2}{2m}+V(x)+\beta\frac{p^4}{3m}+\beta^2\frac{7p^6}{18m}+{\cal
O}(\beta^3),\label{gupHamil}
\end{eqnarray}
which agrees with perturbative version of the KMM proposal to ${\cal
O}(\beta)$ \cite{main}. In the quantum domain, this Hamiltonian
results in the following generalized Schr\"odinger equation in
position space representation:
\begin{eqnarray}
-\frac{\hbar^2}{2m}\frac{\partial^2\psi(x)}{\partial
x^2}+\frac{\beta}{3m}\frac{\partial^4\psi(x)}{\partial
x^4}-\frac{7\beta^2}{18m}\frac{\partial^6\psi(x)}{\partial
x^6}+{\cal O}(\beta^3)+V(x)\,\psi(x)=E\,\psi(x),\label{schro}
\end{eqnarray}
where the extra terms are due to the GUP-corrected terms in
Eq.~(\ref{gupHamil}). As mentioned before, this representation is
naturally perturbative that is apparent from Eq.~(\ref{schro}).

Note that for an operator $A$ which is ``formally'' self-adjoint
($A=A^{\dagger}$) such as (\ref{PP0}) and (\ref{PP2}), this does not
prove that $A$ is truly self-adjoint because in general the domains
${\cal D}(A)$ and ${\cal D}(A^{\dagger})$ may be different. The
operator $A$ with dense domain ${\cal D}(A)$ is said to be
self-adjoint if ${\cal D}(A) ={\cal D}(A^{\dagger})$ and
$A=A^{\dagger}$. For instance, the position operator (\ref{PP0}) is
merely symmetric in this representation, but not self-adjoint. To
see this point, notice that in this representation and in momentum
space, the wave function $\phi(p)$ have to vanish at the boundaries
of the $p$ interval $(-2/3\sqrt{\beta}<p<2/3\sqrt{\beta})$. So $X$
is now the derivative operator $i\hbar\partial/\partial{p}$ on an
interval with Dirichlet boundary conditions. But this means that $X$
cannot be self-adjoint because all candidates for the eigenfunctions
of $X$ (the plane waves) are not in the domain of $X$ because they
do not obey the Dirichlet boundary conditions. Calculating the
domain of the adjoint of $X$ shows that it is larger than that of
$X$, so $X$ is not a true self-adjoint operator, i.e.,
\begin{eqnarray}
\int_{-\frac{2}{3\sqrt{\beta}}}^{+\frac{2}{3\sqrt{\beta}}}\mathrm{d}p\,\psi^{*}(p)\left(i\hbar\frac{\partial}{\partial
p}\right)\phi(p)=\int_{-\frac{2}{3\sqrt{\beta}}}^{+\frac{2}{3\sqrt{\beta}}}\mathrm{d}p\,\left(i\hbar\frac{\partial\psi(p)}{\partial
p}\right)^{*}\phi(p)
\,\,+i\hbar\,\psi^*(p)\phi(p)\Bigg|_{p=+\frac{2}{3\sqrt{\beta}}}\hspace{-.5cm}-i\hbar\,\psi^*(p)\phi(p)\Bigg|_{p=-\frac{2}{3\sqrt{\beta}}}\hspace{-.75cm}.
\end{eqnarray}
Now since $\phi(p)$ vanishes at
$p=\displaystyle\pm\frac{2}{3\sqrt{\beta}}$, $\psi^*(p)$ can take
any arbitrary value there. Therefore, although its adjoint
$X^{\dagger}=i\hbar\partial/\partial p$ has the same formal
expression, it acts on a different space of functions, namely
\begin{eqnarray}
{\cal D}(X)&=&\bigg\{\phi,\phi'\in{\cal
L}^2\left(\frac{-2}{3\sqrt{\beta}},\frac{+2}{3\sqrt{\beta}}\right)\,;
\phi\left(\frac{+2}{3\sqrt{\beta}}\right)=\phi\left(\frac{-2}{3\sqrt{\beta}}\right)=0\bigg\},\\
{\cal D}(X^{\dagger})&=&\bigg\{\psi,\psi'\in{\cal
L}^2\left(\frac{-2}{3\sqrt{\beta}},\frac{+2}{3\sqrt{\beta}}\right)\,;\mbox{no
other restriction on }\psi\bigg\}.
\end{eqnarray}

To better clarify this point, we can also use the the von Neumann's
theorem \cite{Akhiezer,Bonneau}. Thus, we need to find the wave
functions that satisfy the eigenvalue equation
\begin{eqnarray}
X^{\dagger}\phi_{\pm}(p)=i\hbar\partial_p\phi_{\pm}(p)=\pm i\lambda
\phi_{\pm}(x).
\end{eqnarray}
The solutions are
\begin{eqnarray}
\phi_{\pm}(p)=\mathcal{C}_{\pm}e^{\mp \lambda p}.
\end{eqnarray}
Since both $\phi_{\pm}(p)$ belong to ${\cal
L}^2\left(\frac{-2}{3\sqrt{\beta}},\frac{+2}{3\sqrt{\beta}}\right)$,
the deficiency indices are $(1,1)$. Therefore, the position operator
is not self-adjoint but has a one-parameter family of self-adjoint
extensions which is in agreement with the previous result.

\subsection{Free particle}
In ordinary quantum mechanics, the free particle wave function
$u_p(x)$ is defined as the eigenfunction of the momentum operator,
namely $\hat{P}u_p(x)=p\,u_p(x)$ where $p$ is the eigenvalue. Since
the momentum operator in position space is given by
$\hat{P}=-i\hbar\frac{\partial}{\partial x}$, we have
$-i\hbar\frac{\partial u_p(x)}{\partial x}=p\, u_p(x)$ which has the
following solution
\begin{eqnarray}\label{solp}
u_p(x)=\frac{1}{\sqrt{2\pi\hbar}}\exp\left({\frac{ip
x}{\hbar}}\right),
\end{eqnarray}
where the constant of integration is chosen to satisfy
\begin{eqnarray}\label{norm}
\int^{\infty}_{-\infty}u^{\ast}_p(x)u_p(x')\mathrm{d}p=\delta(x-x').
\end{eqnarray}
In the GUP scenario, to find the momentum eigenfunction in position
space, we write the eigenvalue equation as
\begin{eqnarray}\label{p2}
\frac{1-i\sqrt{3}+(-2\beta)^{1/3}\left(-3i\hbar\partial_x+\sqrt{-9\hbar^2\partial^2_x-4/\beta}\right)^{2/3}}{(2\beta)^{2/3}
\left(-3i\hbar\partial_x+\sqrt{-9\hbar^2\partial^2_x-4/\beta}\right)^{1/3}}u_\wp(x)=\wp\,u_\wp(x),
\end{eqnarray}
where $\wp$ is the eigenvalue of $P$. Now, let us take the solution
in the form of Eq.~(\ref{solp})
\begin{eqnarray}
u_\wp(x)=\mathcal{A}\exp\left({\frac{ip\,x}{\hbar}}\right),
\end{eqnarray}
where $p=f(\wp)$. Inserting this solution in Eq.~(\ref{p2}) results
in
\begin{eqnarray}
\frac{1-i\sqrt{3}+(-2\beta)^{1/3}\left(3p+\sqrt{9p^2-4/\beta}\right)^{2/3}}{(2\beta)^{2/3}\left(3p+\sqrt{9p^2-4/\beta}\right)^{1/3}}=\wp,
\end{eqnarray}
or
\begin{eqnarray}\label{p'}
p=\wp-\frac{\beta}{3}\wp^3,
\end{eqnarray}
so we have
\begin{eqnarray}\label{sai}
u_\wp(x)=\mathcal{A}\exp\left[\frac{i}{\hbar}\left(\wp-\frac{\beta}{3}\wp^3\right)x\right].
\end{eqnarray}
The eigenfunctions are normalizable
\begin{eqnarray}
1=\mathcal{A}\mathcal{A}^*\int_{-\frac{2}{3\sqrt{\beta}}}^{+\frac{2}{3\sqrt{\beta}}}\mathrm{d}p
=\frac{4\mathcal{A}\mathcal{A}^*}{3\sqrt{\beta}}.
\end{eqnarray}
Therefore
\begin{eqnarray}\label{sol-n}
u_\wp(x)=\frac{\sqrt{3\sqrt{\beta}}}{2}\exp\left[\frac{i}{\hbar}\left(\wp-\frac{\beta}{3}\wp^3\right)x\right].
\end{eqnarray}
The momentum eigenfunctions now satisfy
\begin{eqnarray}\label{norm2}
\int_{-\frac{2}{3\sqrt{\beta}}}^{+\frac{2}{3\sqrt{\beta}}}u_{\wp}^*(x')u_{\wp}(x)\mathrm{d}p&=&\int_{-1/\sqrt{\beta}}^{+1/\sqrt{\beta}}
\left(1-\beta
\wp^2\right)u_\wp^*(x')u_\wp(x)\mathrm{d}\wp,\\
&=&\frac{3\hbar\sqrt{\beta}}{2(x-x')}\sin\left(\frac{2(x-x')}{3\hbar\sqrt{\beta}}\right).
\end{eqnarray}
Finally, since $\wp_{max}=1/\sqrt{\beta}$, the energy of the free
particle $E=\frac{\wp^2}{2m}$ is bounded from above
\begin{eqnarray}
E_{max}=\frac{1}{2m\beta}.
\end{eqnarray}

To find Eq.~(\ref{sol-n}) we supposed that the coefficient ${\cal
A}$ does not depend on the momentum. If we relax this assumption,
the maximally localized states can be used to find the quasiposition
wave function of the momentum eigenstate $\phi_\wp(p)=\delta(p-\wp)$
in a straightforward way. So inserting $\phi_\wp(p)$ in
Eq.~(\ref{psiQP}) results in
\begin{eqnarray}\label{QP-sol}
\psi_{QP}(\xi)=\sqrt{\frac{3\sqrt{\beta}}{2}} \left(1-\beta
\wp^2\right)
\,\cos\left[\frac{3\pi\sqrt{\beta}}{4}\left(\wp-\frac{\beta}{3}\wp^3\right)\right]
\,\exp\left[\frac{i\xi}{\hbar}\,\left(\wp-\frac{\beta}{3}\wp^3\right)\right]
,
\end{eqnarray}
and therefore ${\cal
A}(\wp)=\sqrt{\frac{3\sqrt{\beta}}{2}}\left(1-\beta \wp^2\right)
\,\cos\left[\frac{3\pi\sqrt{\beta}}{4}\left(\wp-\frac{\beta}{3}\wp^3\right)\right]$.
However, for this case the solutions are no longer the
eigenfunctions of the position operator which is the consequence of
non-self-adjointness property of the position operator. Thus, in
comparison, Eq.~(\ref{QP-sol}) represents the physically acceptable
solutions.

\subsection{Particle in a box}
As another application, let us consider a particle with mass $m$
confined in an infinite one-dimensional box with length $L$
\begin{eqnarray}\label{pot}
V(x)=\left\{
\begin{array}{ll}
0 \hspace{1cm} \,\,0<x<L,\\ \\\infty \hspace{1cm}\mbox{elsewhere} .
\end{array}
\right.
\end{eqnarray}
The corresponding eigenfunctions should satisfy the following
generalized Schr\"odinger equation
\begin{eqnarray}\label{H2}
-\frac{\hbar^2}{2m}\frac{\partial^2\psi_n(x)}{\partial
x^2}+\frac{\beta\hbar^4}{3m}\frac{\partial^4\psi(x)}{\partial
x^4}-\frac{7\beta^2\hbar^6}{18m}\frac{\partial^6\psi(x)}{\partial
x^6}+{\cal O}(\beta^3)=E_n\,\psi_n(x),
\end{eqnarray}
for $0<x<L$ and they also meet the boundary  conditions
$\psi_n(0)=\psi_n(L)=0$. In Refs.~\cite{106,p2}, the above equation
is thoroughly solved to ${\cal{O}}(\beta)$ and its exact eigenvalues
and eigenfunctions are found. Because of the boundary conditions, if
we take the normalized ansatz
\begin{eqnarray}
\psi_n(x)=\sqrt{\frac{\displaystyle 2}{\displaystyle
L}}\sin\left(\frac{\displaystyle n\pi x}{\displaystyle L}\right),
\end{eqnarray}
 Eq.~(\ref{H2}) is satisfied
and we obtain
\begin{eqnarray}
H\psi_n(x)= \left(\varepsilon_n+\frac{4}{3}\beta
m\varepsilon_n^2+\frac{28}{9}\beta^2
m^2\varepsilon_n^3+\frac{80}{9}\beta^3
m^3\varepsilon_n^4+\cdots\right)\psi_n(x)\label{H4}
\end{eqnarray}
where $\varepsilon_n=\frac{\displaystyle
n^2\pi^2\hbar^2}{\displaystyle 2mL^2}$. Now the comparison between
Eqs.~(\ref{H2}) and (\ref{H4}) shows
\begin{eqnarray}\label{EBox}
E_n&=&\varepsilon_n+\frac{4}{3}\beta
m\varepsilon_n^2+\frac{28}{9}\beta^2
m^2\varepsilon_n^3+\frac{80}{9}\beta^3 m^3\varepsilon_n^4+\cdots,\\
&=&\varepsilon_n\left[
\frac{1-i\sqrt{3}+(-2)^{1/3}\left(3\gamma_n+\sqrt{9\gamma_n^2-4}\right)^{2/3}}{4^{1/3}\left(3\gamma_n+\sqrt{9\gamma_n^2-4}\right)^{1/3}}\right]^2,\hspace{.75cm}
\end{eqnarray}
where $\gamma_n=2\beta m \varepsilon_n$. Therefore, to first order
of GUP parameter we have $E_n=\frac{\displaystyle
n^2\pi^2\hbar^2}{\displaystyle 2 mL^2}+\beta\frac{\displaystyle
n^4\pi^4\hbar^4}{\displaystyle 3mL^4}$ which is in agreement with
the result of Ref.~\cite{106}.  These results show that in this GUP
scenario there is no change in the particle in a box eigenfunctions
but there is a positive shift in the energy spectrum which is
proportional to the powers of $\beta$.

We now estimate the energy spectrum using the semiclassical
scheme. For the particle in a box, the Wilson-Sommerfeld
formula
\begin{eqnarray}
\oint p\,\mathrm{d}x= nh,\hspace{1cm}n=1,2,\ldots,
\end{eqnarray}
results in
\begin{eqnarray}
p_n=\frac{nh}{L}.
\end{eqnarray}
Since the high energy momentum $P$ depends on the low energy
momentum through $p_n=P_n-(1/3)\beta P^3_n$ (\ref{Pp}), the
semiclassical energy spectrum is given by
\begin{eqnarray}
E_n^{(SC)}&=&\frac{P_n^2}{2m},\nonumber\\
&=&\left[\frac{1-i\sqrt{3}+(-2\beta)^{1/3}\left(3p_n+\sqrt{9p_n^2-4/\beta}\right)^{2/3}}
{\sqrt{2m}(2\beta)^{2/3}\left(3p_n+\sqrt{9p_n^2-4/\beta}\right)^{1/3}}\right]^2\hspace{-.3cm}.\hspace{.75cm}\label{SME}
\end{eqnarray}
It is straightforward to check that the semiclassical result
(\ref{SME}) exactly coincide with the quantum mechanical spectrum
(\ref{EBox}). Therefore, the number of states is finite
\begin{eqnarray}\label{nmax}
n_{max}=\left\lfloor\frac{2L}{3h\sqrt{\beta}}\right\rfloor,
\end{eqnarray}
where $\lfloor x\rfloor$ denotes the largest integer not greater
than $x$, and the maximal energy of the particle in a box reads
\begin{eqnarray}
E_{max}=\frac{1}{2m\beta}.
\end{eqnarray}
So we found that this upper bound is similar to the case of the free
particle. However, note that because of the presence of the maximum
momentum $P_{max}$ this result is not surprising. Indeed for both
cases we have $E_{max}=P_{max}^2/2m$. Moreover, for the case of the
harmonic oscillator, the maximal semiclassical energy is
$E_{max}^{(SC)}=1/m\beta$ \cite{pedramPLBm}. This value can be
roughly estimated if we associate the same amount of energy to both
kinetic and potential parts of the Hamiltonian, namely
$E_{max}^{(SC)}=E_{max}^{(K)}+E_{max}^{(P)}=2E_{max}$.

It is now worth mentioning that the existence of the upper bound on
the energy spectrum in the GUP scenario is also addressed by Quesne
and Tkachuk in the context of Lorentz-covariant deformed algebra
with minimal length when it is applied to the $(1 + 1)$-dimensional
Dirac oscillator \cite{Quesne}. For that case the energy spectrum
reads
\begin{eqnarray}
|E_{n}|=\frac{c}{\sqrt{\beta}}\sqrt{1+\frac{\beta
m^2c^2-1}{\left(1+\beta m\hbar\omega
n\right)^2}},\hspace{2cm}n=0,1,2,...,
\end{eqnarray}
where $m$ and $\omega$ are the oscillator's mass and frequency,
respectively. Therefore both the deformation parameter and the
energy spectrum are bounded from above, i.e.,
\begin{eqnarray}
|E|_{max}=\frac{c}{\sqrt{\beta}},\hspace{2cm}\beta<\frac{1}{m^2c^2}.
\end{eqnarray}
In comparison, unlike the particle in a box (\ref{nmax}), $n$ is not
bounded and ranges from zero to infinity. However, there is no
restriction on $\beta$ in our formulation in contrary to the
covariant version of the KMM algebra.

\subsection{WKB approximation}
To check the validity of the Wilson-Sommerfeld quantization rule for
this modified quantum mechanics, we need to show that the
zeroth-order wave function, which satisfies the generalized
Schr\"odinger equation (\ref{schro}), can be written as
$\psi(x)\simeq\exp\left[{(i/\hbar)\int p\,\mathrm{d}x}\right]$. So
let us take
\begin{eqnarray}
\psi(x)=e^{i\varphi(x)},
\end{eqnarray}
where $\varphi(x)$ can be expanded as a power series in $\hbar$ in
the semiclassical approximation, i.e.,
\begin{eqnarray}
\varphi(x)=\frac{1}{\hbar}\sum_{n=0}^{\infty}\hbar^n\varphi_n(x).
\end{eqnarray}
So we have
\begin{eqnarray}
\frac{\partial^2\psi(x)}{\partial
x^2}&=&\left(-\varphi'^2+i\varphi''\right)\psi(x),\\
\frac{\partial^4\psi(x)}{\partial
x^4}&=&\big(\varphi'^4-6i\varphi'^2\varphi''-3\varphi''^2-4\varphi'''\varphi'+i\varphi''''\big)\psi(x),\\
\vdots&&\hspace{2cm}\vdots\nonumber
\end{eqnarray}
where the prime indicates the derivative with respect to $x$. Now to
zeroth-order $\varphi(x)\simeq\varphi_0(x)/\hbar$ and for
$\hbar\rightarrow0$ we obtain
\begin{eqnarray}
\varphi_0'^2+\frac{2}{3}\beta\varphi_0'^4+\frac{7}{9}\beta^2\varphi_0'^6+\mathcal{O}(\beta^3)=2m\left(E-V(x)\right).
\end{eqnarray}
Thus, the comparison with Eq.~(\ref{gupHamil}) shows $\varphi_0'=p$
and consequently
\begin{eqnarray}
\psi(x)\simeq \exp\left[{\frac{i}{\hbar}\int p\,\mathrm{d}x}\right],
\end{eqnarray}
which is the usual zeroth-order WKB wave function obeying the
Wilson-Sommerfeld quantization rule.

\section{Generalization to $D$ dimensions}
We now extend the developed formalism in previous sections to $D$
spatial dimensions. We then present the generalized Poisson brackets
in the classical limit and study the density of states.

\subsection{Generalized Heisenberg algebra for $D$ dimensions}
A natural generalization of the one-dimensional commutation relation
(\ref{guph}) that preserves the rotational symmetry is
\begin{eqnarray}\label{com1}
[X_i,P_j]=\frac{i\hbar\delta_{ij}}{1-\beta P^2},
\end{eqnarray}
where $P^2=\sum_{i=1}^{D}P_iP_i$. This relation implies a nonzero
minimal uncertainty and a maximal observable momentum in each
position coordinate. If the components of the momentum operator are
assumed to be commutative
\begin{eqnarray}\label{com2}
[P_i,P_j]=0,
\end{eqnarray}
then the Jacobi identity determines the commutation relations
between the components of the position operator as
\begin{eqnarray}\label{com3}
[X_i,X_j]=\frac{2i\hbar\beta}{\left(1-\beta
P^2\right)^2}\left(P_iX_j-P_jX_i\right),
\end{eqnarray}
which results in a noncommutative geometric generalization of
position space. To exactly satisfy these commutation relations, the
position and momentum operators in the momentum space representation
can be written as
\begin{eqnarray}
P_i \phi(p)&=& p_i\phi(p),\\
X_i\phi(p)&=& \frac{i\hbar}{1 - \beta p^2}\partial_{p_i}\phi(p).
\end{eqnarray}
$X_i$ and $P_j$ are now symmetric operator on the domain
$S_{\infty}$ with respect to the scalar product:
\begin{eqnarray}
\langle\psi|\phi\rangle=\int_{-1/\sqrt{\beta}}^{+1/\sqrt{\beta}}\mathrm{d^D}p
\left(1-\beta p^2\right)\psi^{*}(p)\phi(p),
\end{eqnarray}
where $p^2=\sum_{i=1}^{D}p_ip_i$. The identity operator is
\begin{eqnarray}
1=\int_{-1/\sqrt{\beta}}^{+1/\sqrt{\beta}}\frac{\mathrm{d^D}p}{\left(1-\beta
p^2\right)}|p\rangle\langle p|,
\end{eqnarray}
and the scalar product of momentum eigenstates is
\begin{eqnarray}
\langle p|p'\rangle&=& \frac{\delta^D(p-p')}{1-\beta p^2}.
\end{eqnarray}
In this representation, the components of the momentum operator are
still essentially self-adjoint, however the components of the
position operators are merely symmetric and do not have physical
eigenstates.

Since the commutation relations (\ref{com1})--(\ref{com3}) do not
break the rotational symmetry, we can express the generators of
rotations in terms of the position and momentum operators as
\begin{eqnarray}
L_{ij}\equiv\left(1-\beta P^2\right)\left(X_iP_j-X_jP_i\right),
\end{eqnarray}
as the generalization of the ordinary orbital angular momentum. Now
the momentum space representation of the generators of rotations is
\begin{eqnarray}
L_{ij}\psi(p)=-i\hbar\left(p_i\partial_{p_j}-p_j\partial_{p_i}\right)\psi(p),
\end{eqnarray}
and
\begin{eqnarray}
[P_i,L_{jk}]=i\hbar\left(\delta_{ik}P_j-\delta_{ij}P_k\right),
\end{eqnarray}
\begin{eqnarray}
[X_i,L_{jk}]=i\hbar\left(\delta_{ik}X_j-\delta_{ij}X_k\right),
\end{eqnarray}
\begin{eqnarray}
\hspace{-.5cm}[L_{ij},L_{kl}]=i\hbar\left(\delta_{ik}L_{jl}-\delta_{il}L_{jk}+\delta_{jl}L_{ik}-\delta_{jk}L_{il}\right),
\end{eqnarray}
as well as in ordinary quantum mechanics. However, the geometry is
noncommutative, namely
\begin{eqnarray}
[X_i,X_j]=\frac{-2i\hbar\beta}{\left(1-\beta P^2\right)^2}L_{ij}.
\end{eqnarray}


\subsection{Density of states}
The right hand side of Eq.~(\ref{guph}) shows that the ``effective''
value of $\hbar$ is $P$ dependent. So the size of the unit cell in
the phase space that is occupied by each quantum state can be also
considered of as being momentum dependent. This fact changes the
momentum dependence of the density of states and affects the
calculation of cosmological constant, blackbody radiation spectrum,
etc. Similar to the KMM algebra \cite{prd}, we should check that any
volume of the phase space evolves such that the number of states
inside it does not change with respect to time as the analog of the
Liouville theorem.

The Poisson brackets in classical mechanics correspond quantum
mechanical commutators via
\begin{equation}
\frac{1}{i\hbar} [A,B] \Longrightarrow \{A,B\}.
\end{equation}
Thus the classical limits of Eqs.~(\ref{com1})--(\ref{com3}) are
given by
\begin{eqnarray}
\{X_i,P_j\} & = & \frac{\delta_{ij}}{ 1 - \beta P^2 },\\
\{P_i,P_j\} & = & 0,\\
\{X_i,X_j\} & = & \frac{2\beta}{\left(1-\beta
P^2\right)^2}\left(P_iX_j-P_jX_i\right),
\end{eqnarray}
and the Heisenberg equations for the coordinates and momenta read ($i,j$ run over the spatial
dimensions and the summation convention is assumed)
\begin{eqnarray}\label{xdot}
\dot{X}_i & = & \{X_i,H\}=\{X_i,P_j\}\frac{\partial H}{\partial P_j}
    + \{X_i,X_j\}\frac{\partial H}{\partial X_j},\\
\dot{P}_i & = & \{P_i,H\} = -\{X_j,P_i\}\frac{\partial H}{\partial
X_j}.
\end{eqnarray}
Note that in one dimension Eq.~(\ref{xdot}) implies that although
the momentum is bounded from above, the velocity
\begin{equation}
\dot{X}=\{X,H\}=\frac{P}{m\left(1-\beta P^2\right)},
\end{equation}
ranges from $-\infty$ to $+\infty$ as $P$ goes to
$\pm\frac{1}{\sqrt{\beta}}$. We now prove that the weighted phase
space volume
\begin{equation}
\Bigl( 1-\beta P^2 \Bigr)^{D}
\mathrm{d}^DX\,\mathrm{d}^DP,\label{hajm}
\end{equation}
is invariant under time evolution as the analog of the Liouville
theorem. The evolution of $X_i$ and $P_i$ during an infinitesimal
time interval $\delta t$ is
\begin{eqnarray}
X_i' & = & X_i + \delta X_i,\\
P_i' & = & P_i + \delta P_i,
\end{eqnarray}
where
\begin{eqnarray}
\delta X_i & = & \left[ \{X_i,P_j\}\,\frac{\partial H}{\partial P_j}
           + \{X_i,X_j\}\,\frac{\partial H}{\partial X_j}
      \right]\delta t,\\
\delta P_i & = &  -\{X_j,P_i\}\,\frac{\partial H}{\partial
X_j}\delta t.
\end{eqnarray}
After this infinitesimal evolution, the infinitesimal phase space
volume is changed according to
\begin{equation}
\mathrm{d}^DX'\,\mathrm{d}^DP' = \left|
\dfrac{\partial(X'_1,\cdots,X'_D,P'_1,\cdots,P'_D)}
               {\partial(X_1, \cdots,X_D, P_1, \cdots,P_D)}
  \right|
\mathrm{d}^DX\,\mathrm{d}^DP.
\end{equation}
where
\begin{equation}
\begin{array}{ll}
\displaystyle\frac{\partial X'_i}{\partial X_j} = \delta_{ij} +
\displaystyle\frac{\partial \delta X_i}{\partial X_j}, &\qquad
\displaystyle\frac{\partial X'_i}{\partial P_j} =
\displaystyle\frac{\partial \delta X_i}{\partial P_j},\\
\displaystyle\frac{\partial P'_i}{\partial X_j} =
\displaystyle\frac{\partial \delta P_i}{\partial X_j}, &\qquad
\displaystyle\frac{\partial P'_i}{\partial P_j} = \delta_{ij} +
\displaystyle\frac{\partial \delta P_i}{\partial P_j}.
\end{array}
\end{equation}
The Jacobian can be calculated to first-order in $\delta t$ as
\begin{equation}
\left| \dfrac{\partial(X'_1,\cdots,X'_D,P'_1,\cdots,P'_D)}
               {\partial(X_1, \cdots,X_D, P_1, \cdots,P_D)}
\right| = 1 + \left( \frac{\partial\delta X_i}{\partial X_i}
         + \frac{\partial\delta P_i}{\partial P_i}
    \right)
  + \cdots \;.
\end{equation}
So we have
\begin{eqnarray}
\left(\frac{\partial\delta X_i}{\partial X_i}
    + \frac{\partial\delta P_i}{\partial P_i}\right)\frac{1}{\delta t}&=& \frac{\partial}{\partial X_i}
      \left[ \{X_i,P_j\}\,\frac{\partial H}{\partial P_j}
           + \{X_i,X_j\}\,\frac{\partial H}{\partial X_j}
      \right]
    - \frac{\partial}{\partial P_i}
      \left[ \{X_j,P_i\}\,\frac{\partial H}{\partial X_j}
      \right], \nonumber\\
&=& \left[ \frac{\partial}{\partial X_i}\{X_i,P_j\} \right]
      \frac{\partial H}{\partial P_j}
    + \{X_i,P_j\}\frac{\partial^2 H}{\partial X_i \partial P_j}
    + \left[ \frac{\partial}{\partial X_i}\{X_i,X_j\} \right]
      \frac{\partial H}{\partial X_j} \nonumber\\
    &&+ \{X_i,X_j\}\frac{\partial^2 H}{\partial X_i
\partial X_j}- \left[ \frac{\partial}{\partial P_i}\{X_j,P_i\} \right]
      \frac{\partial H}{\partial X_j}
    - \{X_j,P_i\}\frac{\partial^2 H}{\partial P_j \partial X_i}, \nonumber\\
&=&  \left[ \frac{\partial}{\partial X_i}\{X_i,X_j\} \right]
      \frac{\partial H}{\partial X_j}
    - \left[ \frac{\partial}{\partial P_i}\{X_j,P_i\} \right]
      \frac{\partial H}{\partial X_j}, \nonumber\\
&=& \Biggl[ -\frac{ 2\beta (D-1)}{\left(1-\beta P^2\right)^2 } \,P_j
      \Biggr] \frac{\partial H}{\partial X_j}
     -\Biggl[ \frac{ 2\beta}{\left(1-\beta
P^2\right)^2 }\,P_j
      \Biggr]\frac{\partial H}{\partial X_j}, \nonumber\\
&=&  \frac{ -2\beta D}{\left(1-\beta P^2\right)^2} P_j
\frac{\partial H}{\partial X_j},
\end{eqnarray}
which to first-order in $\delta t$ results in
\begin{equation}
\mathrm{d}^DX'\,\mathrm{d}^DP' = \mathrm{d}^DX\,\mathrm{d}^DP
  \left[ 1
     - \frac{ 2\beta D}{\left(1-\beta P^2\right)^2} P_j
\frac{\partial H}{\partial X_j}\delta t
  \right].\label{dx}
\end{equation}
Moreover
\begin{eqnarray}
1 - \beta {P'}^2 & = & 1 - \beta (P_i + \delta P_i)^2,\nonumber\\
 & = & 1 - \beta\left( P^2 + 2P_i \delta P_i + \cdots
 \right),\nonumber\\
 & = & 1 - \beta\left( P^2
                  - 2P_i\{X_i,P_j\}\frac{\partial H}{\partial X_j}\delta t
                  + \cdots
                \right),\nonumber\\
& = & 1 - \beta\left( P^2
                - \frac{2 P_i}{1-\beta P^2}\frac{\partial H}{\partial X_i}\delta t
                + \cdots
               \right),\nonumber\\
& = & (1-\beta P^2)
       + \frac{2 \beta P_i}{1-\beta P^2}\frac{\partial H}{\partial X_i}\delta t +
       \cdots, \nonumber\\
& = & (1-\beta P^2)
      \left[ 1 + \frac{2 \beta P_i}{\left(1-\beta P^2\right)^2}\frac{\partial H}{\partial X_i}\delta t + \cdots
      \right].
\end{eqnarray}
Therefore, to first-order in $\delta t$
\begin{eqnarray}
\hspace{-1cm}\left(1 - \beta {P'}^2\right)^D=\left(1 - \beta
{P}^2\right)^D\left[ 1 + \frac{2 \beta D }{\left(1-\beta
P^2\right)^2}P_i\frac{\partial H}{\partial X_i}\delta t
\right],\hspace{.5cm}\label{wei}
\end{eqnarray}
Now using Eqs.~(\ref{dx}) and (\ref{wei}), it is obvious that the
weighted phase space volume Eq.~(\ref{hajm}) is an invariant, i.e.,
\begin{equation}
\left(1 - \beta {P'}^2\right)^D\mathrm{d}^DX'\,\mathrm{d}^DP'
=\left(1 - \beta {P}^2\right)^D \mathrm{d}^DX\,\mathrm{d}^DP.
\end{equation}

\subsection{The cosmological constant}
The cosmological constant can be obtained by summing over the
zero-point energies of the harmonic oscillator's momentum states.
Using the canonical form of the zero-point energy of each oscillator
with mass $m$
\begin{eqnarray}
\frac{1}{2}\hbar\omega=\frac{1}{2}\sqrt{p^2+m^2},
\end{eqnarray}
the sum over all momentum states per unit volume is
\begin{eqnarray}
\Lambda(m)&=&\int \mathrm{d}^3p \left(1-\beta
p^2\right)^3\left(\frac{1}{2}\sqrt{p^2+m^2}\right),\nonumber\\
&=&2\pi\int_0^{1/\sqrt{\beta}}\mathrm{d}p\left(1-\beta
p^2\right)^3p^2\sqrt{p^2+m^2},\nonumber\\
&=&\frac{\pi}{20\,{\beta }^2}f(\beta m^2),
\end{eqnarray}
where
\begin{eqnarray}
f(x)=\frac{1}{96}&\Big[&( 96 +  192x+476x^2+380x^3+105x^4)\sqrt{1 + x}\nonumber\\
&&-(480x^2+720x^3 + 450x^4 + 105x^5) \cosh^{-1}\left(\sqrt{x
}\right)\Big],
\end{eqnarray}
and $f(0)=1$. In the massless limit we find
\begin{eqnarray}
\Lambda(0)=\frac{\pi}{20\beta^2}=\frac{1}{10}\left[\Lambda(0)\right]^{\mathrm{KMM}},
\end{eqnarray}
that is ten times smaller than the massless cosmological constant
predicted by the KMM proposal \cite{prd}. This finite result is due
to the vanishing of the density of states at high momenta where
$p=1/\sqrt{\beta}$ plays the role of the UV cutoff. So in this
scenario we do not need to put by hand an arbitrary scale as the UV
cutoff and the cosmological constant is automatically rendered
finite. Note that since $1/\sqrt{\beta}$ is proportional to the
Planck mass $M_{Pl}$, $\Lambda(0)$ is too large in practice and
consequently the cosmological constant problem still remains
unsolved. However, our formulation gives the better estimation of
$\Lambda$ with respect to that obtained in the KMM framework.

\subsection{The blackbody radiation spectrum}
\begin{figure}[t]
\begin{center}
\includegraphics[width=8cm]{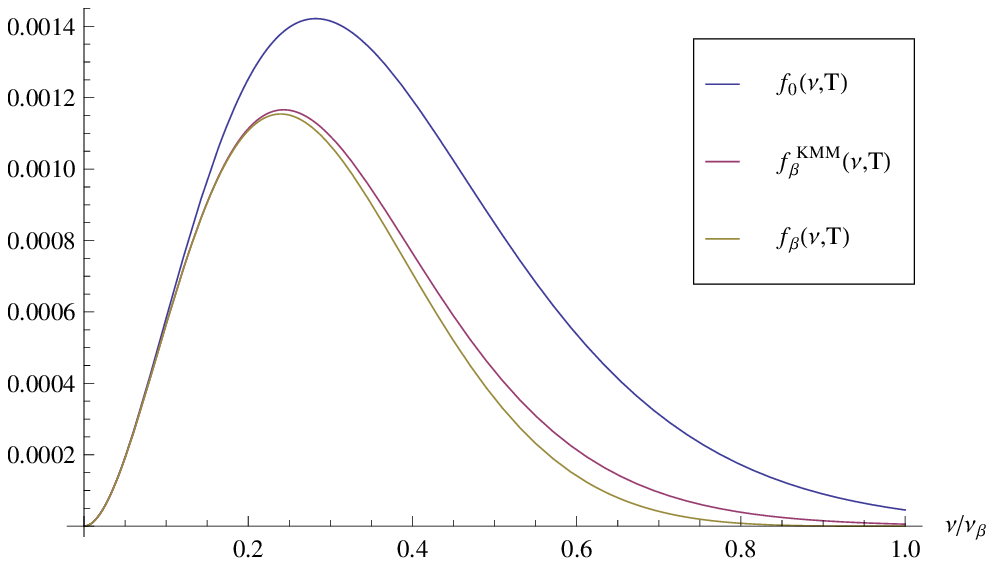}
\caption{\label{fig4}The blackbody radiation spectrum in the GUP
framework at temperature $T=0.1\, T_\beta$.}
\end{center}
\end{figure}

\begin{figure}[t]
\begin{center}
\includegraphics[width=8cm]{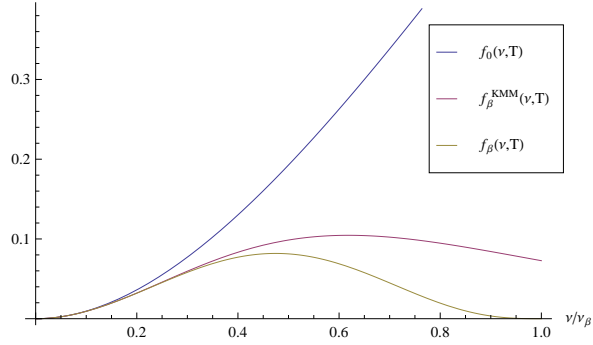}
\caption{\label{fig5}The blackbody radiation spectrum in the GUP
framework at temperature $T=T_\beta$.}
\end{center}
\end{figure}

Because of the weight factor $(1-\beta P^2)^3$ in 3-dimensions, the
average energy of the electromagnetic field per unit volume at
temperature $T$ is given by
\begin{eqnarray}
\langle E\rangle& = & \frac{8\pi}{c^3}\int_0^\infty \mathrm{d}\nu\,
      \left(1-\beta \left(\frac{h\nu}{c}\right)^2\right)^3
      \left(\frac{h\nu^3}{e^{h\nu/k_B T} - 1}\right),\nonumber \\
& = & \int_0^\infty \mathrm{d}\nu\,u_\beta(\nu,T),
\end{eqnarray}
where
\begin{eqnarray}
u_\beta(\nu,T)=\left(1-
\left(\frac{\nu}{\nu_{\beta}}\right)^2\right)^3u_0(\nu,T).
\end{eqnarray}
Here
\begin{eqnarray}
u_0(\nu,T)=\frac{8\pi h \nu^3}{c^3}\frac{1}{e^{h\nu/k_B T}-1},
\end{eqnarray}
is the ordinary spectrum function and $\nu_\beta= c/h\sqrt{\beta}$.
To show the effect of the minimal length uncertainty and the maximal
momentum on the shape of the spectral function, we have depicted the
functions
\begin{eqnarray}
f_0(\nu,T) & = & \frac{(\nu/\nu_{\beta})^3}{ e^{(\nu/\nu_{\beta})(T_{\beta}/T)} - 1 },\\
f_{\beta}(\nu,T) & = & \left(1-(\nu/\nu_{\beta})^2\right)^3
\,f_0(\nu,T),
\end{eqnarray}
in Figs. \ref{fig4} and \ref{fig5}, and compared them with the case
of just the minimal length uncertainty \cite{prd}
\begin{eqnarray}
f_{\beta}^{\mathrm{KMM}}(\nu,T)=
\frac{1}{\left(1+(\nu/\nu_\beta)^2\right)^3}\,f_0(\nu,T),
\end{eqnarray}
where $T_\beta = c/k_B \sqrt{\beta}$. As the figure shows, for small
frequencies ($\nu\ll\nu_\beta$), $f_{\beta}(\nu,T)$ closely
coincides with $f_{\beta}^{\mathrm{KMM}}$. However, it deviates from
$f_{\beta}^{\mathrm{KMM}}$ as the frequency increases.

\section{Conclusions}
In this Letter, we studied a higher order generalized uncertainty
principle that implies both a minimal length uncertainty and a
maximal momentum proportional to $\hbar\sqrt{\beta}$ and
$1/\sqrt{\beta}$, respectively. We found maximally localized states
and presented a formally self-adjoint representation that preserves
the ordinary nature of the position operator and results in the
perturbative generalized Schr\"odinger equation. We exactly solved
the problems of the free particle and the particle in a box and
showed that the existence of the maximal momentum
$P_{max}=1/\sqrt{\beta}$ is manifest through this representation. We
then generalized this proposal to $D$ dimensions and found the
invariant density of states. We showed that the blackbody radiation
spectrum are modified at high frequencies and compared the results
with the KMM proposal. Although the cosmological constant was
rendered finite, the smallness of the GUP parameter resulted in a
large cosmological constant that could not solve the cosmological
constant problem. However, our calculated cosmological constant is a
better estimation with respect to the presence of just the minimal
length.

\section*{Acknowledgements}
I would like to thank Stephane Detournay for introducing his work
about maximally localized states.

\end{document}